\documentclass[12pt]{article}
\usepackage{amsmath}
\usepackage{newtxtext, newtxmath}
\usepackage{graphicx}
\usepackage{geometry}
\usepackage{indentfirst}
\def\vector#1{\mbox{\boldmath $#1$}}
\usepackage{hyperref}

\title{Numerical investigation of internal plasma currents \\ in a magnetic nozzle}
\author{Kazuma Emoto,\textsuperscript{1} Kazunori Takahashi,\textsuperscript{2} and Yoshinori Takao\textsuperscript{3}}
\date{}
\usepackage[style=phys,articletitle=false,biblabel=super,chaptertitle=false,pageranges=false]{biblatex}
\DeclareCiteCommand{\citenum}{}{\printfield[bibhyperref]{labelnumber}}{}{}

\begin{document}

\maketitle

\noindent
\textsuperscript{1}\textit{Department of Mechanical Engineering, Materials Science, and Ocean Engineering, Yokohama National University, Yokohama 240-8501, Japan}

\noindent
\textsuperscript{2}\textit{Department of Electrical Engineering, Tohoku University, Sendai 980-8579, Japan}

\noindent
\textsuperscript{3}\textit{Division of Systems Research, Yokohama National University, Yokohama 240-8501, Japan}

\noindent
The authors to whom correspondence may be addressed: kazuma-emoto-vh@ynu.jp, takao@ynu.ac.jp

\section*{ABSTRACT}

Two-dimensional fully kinetic particle-in-cell simulations of an electrodeless plasma thruster, which uses a magnetic nozzle, were conducted to investigate the thrust generation induced by the internal plasma current. The results clearly show that the $\vector{E} \times \vector{B}$ and diamagnetic current densities are the major components of the internal plasma current. The simulated pressure structures reproduced the experimentally observed structures well. The results for various magnetic field strengths reveal that the $\vector{E} \times \vector{B}$ effect decreases and the diamagnetic effect becomes dominant with an increase in the magnetic field strength; this demonstrates the significant contribution of the diamagnetic effect in thrust generation.

\section{INTRODUCTION}

Electrodeless plasma thrusters have been developed worldwide and are expected to have a long lifetime.\supercite{Takahashi2019_rmpp} They mainly consist of a radiofrequency (rf) antenna, a dielectric tube, and a solenoid. The solenoid generates both an axial magnetic field and a magnetic nozzle downstream of the dielectric tube, which transports the plasma axially along the field lines and accelerates it.\supercite{Sasoh1994_pop, Arefiev2004_pop, Charles2007_psst, Ahedo2010_pop, Terasaka2010_pop, Sekine2020_pop} The physics underlying the acceleration of the magnetic nozzle has also been investigated in astrophysics,\supercite{Revet2021_nature} and the performance of electrodeless plasma thrusters has been improved as an engineering application. In a recent experiment, a thrust of 70 mN was delivered and a thruster efficiency of 20\% for an rf power of 6 kW was achieved.\supercite{Takahashi2021_sr}

In the magnetic nozzle, a Lorentz force, attributed to the azimuthal current and the magnetic field, is exerted on the plasma, which produces thrust in the axial direction. Previous studies implied that the azimuthal current in electrodeless plasma thrusters was mainly the result of a diamagnetic effect, both theoretically and experimentally, when varying the magnetic field strength.\supercite{Takahashi2011_prl, Fruchtman2012_pop, Takahashi2013_prl} In addition, the azimuthal current in the electrodeless plasma thruster was measured experimentally, with the results indicating that the $\vector{E} \times \vector{B}$ drift current decreased with increasing magnetic field strength and the diamagnetic effect became dominant.\supercite{Takahashi2016_psst} The magnetic field gradient has also been demonstrated to significantly affect the electron transport.\supercite{Oudini2019_pop, Li2020_psst} However, the experiment did not accurately evaluate other drifts such as grad-$B$ and curvature drifts.

The diamagnetic effect in the magnetic nozzle was analyzed using a two-fluid model.\supercite{Merino2016_psst} Moreover, particle-in-cell simulations were also conducted to investigate potential structures in the magnetic nozzle.\supercite{Rao2012_pop, Singh2013_pop} These numerical simulations focused on the downstream region of the thruster and did not simulate the plasma source; however, the connection between the two different regions of the plasma source and the magnetic nozzle is the key element that would need to be analyzed to understand the plasma dynamics in an electrodeless plasma thruster that uses a magnetic nozzle. In our previous studies, we conducted particle-in-cell simulations of an electrodeless plasma thruster with Monte Carlo collisions (PIC-MCCs) to investigate the axial momentum loss to the lateral dielectric wall and the effects of neutral distributions. These simulations focused on the source region by employing a closed boundary for the source tube exit.\supercite{Takao2015_pop, Takase2018_pop} However, PIC-MCC simulations that take into account both the plasma source and the magnetic nozzle acceleration have not been performed yet. These simulations would obviate the need for assumptions about the density and velocity distributions for the plasma injection at the magnetic nozzle.

In this study, we conducted PIC-MCC simulations that include both the source and magnetic nozzle regions of the electrodeless plasma thruster and investigated the internal plasma current. Here, PIC-MCC simulations can accurately and self-consistently derive the azimuthal current induced by charged particles in the magnetic nozzle. The $\vector{E} \times \vector{B}$ and diamagnetic drift currents can also be computed from the potential and pressure profiles of the electrons and are compared with the net current profile; the results demonstrate that the internal azimuthal current is diamagnetic in the case of a strong magnetic field. The numerical simulation additionally demonstrated that the Lorentz force induced by the net current and the magnetic nozzle can be enhanced by increasing the magnetic field strength, which is consistent with the previous observation.\supercite{Takahashi2013_prl, Takahashi2016_psst}

In Sec.~\ref{model}, the numerical model employed in this study is briefly described. Sec.~\ref{results_and_discussion} presents the calculation of the net electron current density using the electron motions for three solenoid currents. These results are compared with the $\vector{E} \times \vector{B}$ drift current resulting from the electrostatic field and the diamagnetic drift current due to the electron pressure gradient. As the solenoid current increases, the net electron current density is increasingly dominated by the diamagnetic effect. In addition, the Lorentz force induced by the net electron current density is discussed.

\section{NUMERICAL MODEL}\label{model}

\begin{figure}
    \centering
    \includegraphics{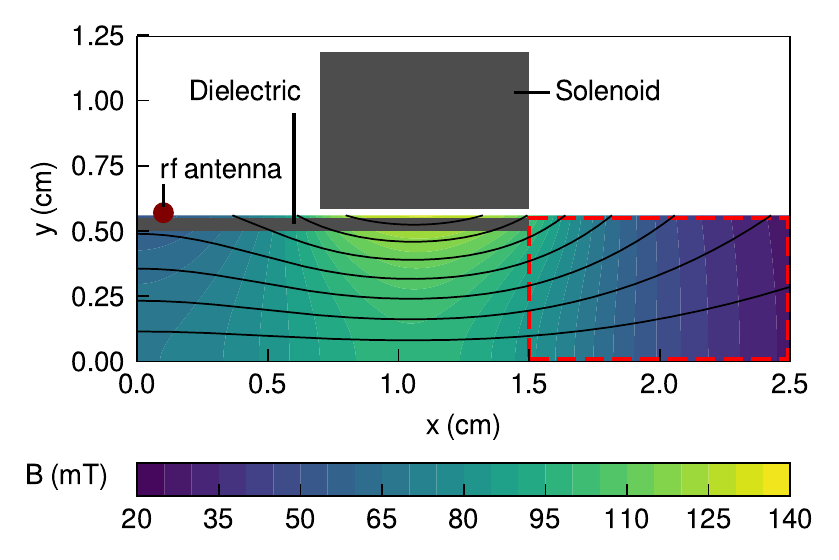}
    \caption{Schematic of the calculation area and the magnetic field strength produced by the solenoid at $I_B$ = 2.0 kA. The solid black curves indicate the magnetic field. Although the calculation was conducted for the entire area of 2.5 cm $\times$ 0.56 cm, the results that are subsequently presented are only those obtained for the area within the dashed red rectangle.}
    \label{fig:model}
\end{figure}

We decided to employ a two-dimensional and symmetric calculation model to avoid singularities on the central axis and to reduce the calculation cost compared with the axisymmetric model. Figure \ref{fig:model} shows a schematic of the area of the electrodeless plasma thruster that was the subject of this study. The calculation model consisted of an rf antenna, a dielectric plate, and a solenoid. The rf antenna supplies power to the plasma, and the dielectric plate confines the plasma to the $y$-direction. The solenoid produces a magnetostatic field that forms a magnetic nozzle that accelerates the plasma. The size of the area included in the calculation is 2.5 cm $\times$ 0.56 cm; this area contains a dielectric with a length of 1.5 cm and a thickness of 0.05 cm. Only the solenoid magnetic field is solved using a calculation area that is ten times larger. It should be noted that the height of the discharge chamber employed in our simulation is 1 cm, which is approximately one-sixth of the diameter of the thruster (6.4 cm in this experiment).\supercite{Takahashi2016_psst} The calculation model is symmetric about the $x$- and $y$-axes such that the plasma is extracted bidirectionally in this simulation. Although the bidirectional thruster differs from the normal unidirectional thruster, the plasma physics in the magnetic nozzle within $x$ = 1.5--2.5 cm would remain essentially unchanged. In addition, bidirectional thrusters have also been proposed and have been demonstrated for space debris removal.\supercite{Takahashi2018_sr} The calculation area is divided into 50 $\mu$m $\times$ 50 $\mu$m cells. In Sec.~\ref{results_and_discussion}, we focus on the downstream region of the thruster to investigate the internal plasma currents in the magnetic nozzle and we present only the results that pertain to the area demarcated by the dashed red rectangle in Fig.~\ref{fig:model}. The simulation results presented in Sec.~\ref{results_and_discussion} are averaged over 30 $\mu$s after reaching the steady state.

\begin{figure}
    \centering
    \includegraphics{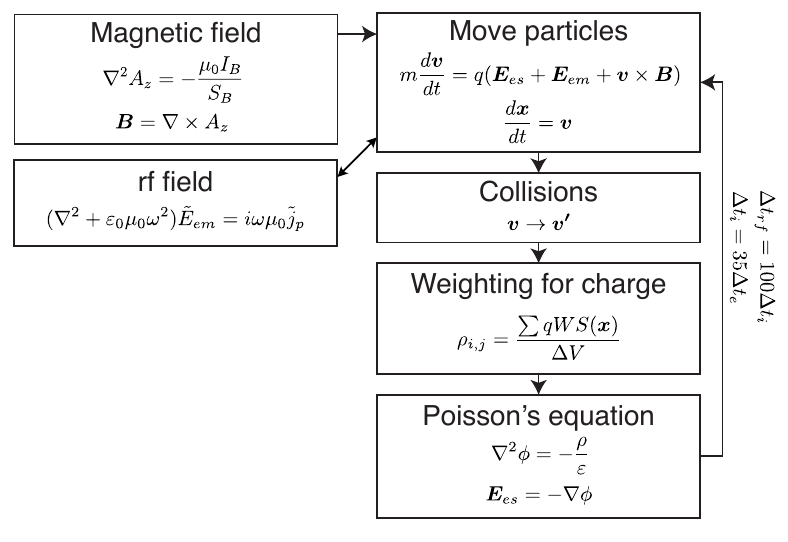}
    \caption{Calculation flow for the PIC-MCC simulation.}
    \label{fig:flow_chart}
\end{figure}

Figure \ref{fig:flow_chart} shows the calculation flow for the PIC-MCC simulation. Our PIC-MCC simulation mainly considers the kinetics of charged particles, electron-neutral collisions, weighting for the charge, and an electrostatic field. Details of the PIC-MCC simulation appear in our previous papers\supercite{Takao2015_pop, Takase2018_pop}; hence, we only include a brief description of our PIC-MCC simulation in this paper.

The equations of motion of the charged particles are written as

\begin{equation}
    \label{eq:equation_of_motion_1}
    m \frac{d\vector{v}}{dt} = q (\vector{E}_{es} + \vector{E}_{em} + \vector{v} \times \vector{B}),
\end{equation}

\begin{equation}
    \label{eq:equation_of_motion_2}
    \frac{d\vector{x}}{dt} = \vector{v},
\end{equation}

\noindent
where $m$, $\vector{v}$, $q$, and $\vector{x}$ are the mass, velocity, charge, and position of the charged particles, respectively. Further, $t$ is the time, $\vector{E}_{es}$ is the electrostatic field produced by the charged particles, $\vector{E}_{em}$ is the electric field induced by the rf antenna, and $\vector{B}$ is the magnetostatic field produced by the solenoid. Eqs.~\eqref{eq:equation_of_motion_1} and \eqref{eq:equation_of_motion_2} are solved using the Boris method.\supercite{Birdsall2004_crc} We chose xenon as the propellant and treat singly charged xenon ions Xe$^+$ and electrons e$^-$ as charged particles in the simulation. In addition, we calculated the initial velocities of ions and electrons using the Maxwell distribution and inserted 100 ions and 100 electrons per cell in the calculation area. A particle that collides with the dielectric is eliminated from the simulation, and its charge is added to the surface charge. The surface charge density $\sigma_s$ is calculated as

\begin{equation}
    \sigma_s = \frac{e (N_{s,i} - N_{s,e})}{\Delta S},
\end{equation}

\noindent
where $e$ is the elementary charge, $N_{s,i}$ and $N_{s,e}$ are the number of ions and electrons in a dielectric cell, respectively, and $\Delta S$ is the cell area.

The potential generated by true charges (charged particles and surface charge) $\phi_t$ satisfies Poisson's equation as follows:

\begin{equation}
    \label{eq:true_poisson_equation}
    \nabla^2 \phi_t = - \frac{e(n_i - n_e) + \sigma_s / \Delta y}{\varepsilon_0},
\end{equation}

\noindent
where $n_i$ and $n_e$ are the ion and electron number densities, respectively, $\Delta y$ is the cell size in the $y$-direction, and $\varepsilon_0$ is the vacuum permittivity. Equation \eqref{eq:true_poisson_equation} is solved using fast Fourier transformation. Then, assuming that the electrostatic field generated by the polarization charge and is perpendicular to the dielectric surface does not exist in the vacuum, the polarization charge density on the dielectric $\sigma_p$ is obtained from

\begin{equation}
    \label{eq:polarization_charge_density}
    \sigma_p = \frac{\chi \varepsilon_0 E_{es,t,\perp}}{\Delta y (1 + \chi)},
\end{equation}

\noindent
where $\chi$ is the susceptibility and $E_{es,t,\perp}$ is the electrostatic field generated by the true charges and is perpendicular to the dielectric surface. In this study, the susceptibility $\chi$ was set to 2.8. Using the polarization charge density $\sigma_p$, the potential generated by the polarization charge $\phi_p$ is obtained as

\begin{equation}
    \label{eq:polarization_poisson_equation}
    \nabla^2 \phi_p = - \frac{\sigma_p}{\varepsilon_0 \Delta y},
\end{equation}

\noindent
using fast Fourier transformation. The total potential $\phi$ is obtained from

\begin{equation}
    \phi = \phi_t + \phi_p,
\end{equation}

\noindent
and the electrostatic field $\vector{E}_{es}$ is obtained from

\begin{equation}
    \label{eq:electrostatic_field}
    \vector{E}_{es} = - \nabla \phi.
\end{equation}

\noindent
The electrostatic field on the dielectric at $y$ = 0.5 cm is obtained from

\begin{equation}
    \nabla \cdot \vector{E}_{es} = \frac{e(n_i - n_e) + (\sigma_s + \sigma_p) / \Delta y}{\varepsilon_0},
\end{equation}

\noindent
such that $\vector{E}_{es}$ satisfies Gauss's law. We set the Dirichlet boundary condition as $\phi$ = 0 at $x$ = 2.5 cm and $y$ = 0.56 cm and the Neumann boundary condition as $\partial \phi/\partial x$ = 0 at $x$ = 0 and $\partial \phi/\partial y$ = 0 at $y$ = 0. Because the Dirichlet boundary condition produces a sheath near the boundaries, the calculation model employed in this study simulates and validates the thruster as though it was operated in a metal vacuum chamber in experiments. However, a sheath was also generated, even in experiments near the chamber wall.\supercite{Takahashi2018_prl} The plasma current away from the sheath is expected to be unaffected. It should be noted that we focus on the phenomena only in a small area downstream of the thruster exit.

The complex electric field induced by the rf antenna $\tilde{E}_{em}$ is written as 

\begin{equation}
    \label{eq:induced_electric_field}
    (\nabla^2 + \varepsilon_0 \mu_0 \omega^2) \tilde{E}_{em} = i \omega \mu_0 \tilde{j}_p,
\end{equation}

\noindent
where $\mu_0$ is the vacuum permeability, $\omega$ is the rf angular frequency, $i$ is the imaginary unit, and $\tilde{j}_p$ is the complex plasma current density.\supercite{Takao2010_jap, Takao2012_jap} We also solve Eq.~\eqref{eq:induced_electric_field} using fast Fourier transformation once in every ten rf periods. The complex plasma current density $\tilde{j}_p$ is obtained as follows:

\begin{equation}
    \tilde{j}_p = \frac{\sum q W v_z}{\Delta V} \exp (i \Delta \psi),
\end{equation}

\noindent
where $W$ is the particle weight, $\Delta V$ is the cell volume, and $\Delta \psi$ is the phase difference between the rf and plasma currents.\supercite{Takao2010_jap} Here, the particle weight is set to 2.5 $\times$ 10$^{6}$. The complex plasma current density $\tilde{j}_p$ is averaged over ten rf periods to reduce the error due to finite particles. The phase difference between $\tilde{E}_{em}$ and $\tilde{j}_p$ is obtained by subtracting the phase of the maximum rf electric field from that of the maximum plasma current. Here, $\tilde{E}_{em}$ on the boundary at $x$ = 1.5 cm is assumed to be zero, and $\tilde{E}_{em}$ on the boundary at $y$ = 0.56 cm is calculated from the plasma current $\tilde{j}_p$ and the rf current using the Biot-Savart law. The absolute value of $\tilde{E}_{em}$ oscillates in the sine function using the rf period and the phase difference.\supercite{Takao2010_jap} Here, the magnetic field induced by the rf antenna is neglected because it is smaller than the magnetic field of the solenoid in the downstream region. The rf current is controlled to maintain the power absorption of the plasma constant at 3.5 W/m and decreases or increases when the total absorbed power is larger or smaller than 3.5 W/m, respectively. The rf frequency was set to 80 MHz. We set the time step of ions to 0.125 ns as 1/100 of the rf period and that of electrons to 3.57 ps as 1/35 of the ion time step.

The solenoid magnetic field $\vector{B}$ is written as 

\begin{equation}
    \label{eq:solenoid_magentic_field}
    \nabla^2 A_z = - \frac{\mu_0 I_B}{S_B},
\end{equation}

\begin{equation}
in    \label{eq:magnetic_field}
    \vector{B} = \nabla \times A_z,
\end{equation}

\noindent
where $A_z$ is the vector potential in the $z$-direction produced by the solenoid, $I_B$ is the solenoid current, and $S_B$ is the cross-sectional area of the solenoid. We also solve Eq.~\eqref{eq:solenoid_magentic_field} using fast Fourier transformation and obtain the magnetic field of the solenoid $\vector{B}$ from Eq. \eqref{eq:magnetic_field}. The solenoid currents were set to 0.1, 0.4, and 2.0 kA to investigate the dependence on the magnetic field strength. Figure \ref{fig:model} also shows the magnetic field strength as a color map and the magnetic field lines as solid black lines at $I_B$ = 2.0 kA. At a strong magnetic field of $I_B$ = 2.0 kA, the typical electron number density $n_e$ and electron temperature $T_e$ are 1 $\times$ 10$^{17}$ m$^{-3}$ and 5 eV, respectively. Under these conditions, the Debye length and the plasma frequency were estimated to be approximately 52.6 $\mu$m and 35.7 ps, respectively. Therefore, the time step of electrons and the cell size in the simulations were smaller than the Debye length and plasma frequency.

We treat elastic, excitation, and ionization collisions as electron-neutral collisions in the simulation using the null-collision method.\supercite{Vahedi1995_cpc} Here, ion-neutral collisions are neglected for simplicity because the mean free path for ion-neutral collisions is estimated to be 5 cm using a cross-section of 1 $\times$ 10$^{-18}$ m$^{-3}$ and a neutral density of 2 $\times$ 10$^{19}$ m$^{-3}$. Many ions do not collide with neutral particles in the calculation area of 2.5 cm $\times$ 0.56 cm. The density of these neutral particles was set to $2 \times 10^{19}$ m$^{-3}$, and the temperature of these particles was set to 300 K.

The PIC simulation can directly provide the internal plasma current density by considering the velocity components of the charged particles. The plasma current determined by the fluid model was discussed in a previous report.\supercite{Takahashi2016_psst} To compare the simulation results with the results of the experiment, the electron pressure $p_e$ is calculated from

\begin{equation}
    \label{eq:electron_pressure}
    p_e = n_e k_B T_e,
\end{equation}

\noindent
where $n_e$ is the electron number density, $k_B$ is the Boltzmann constant, and $T_e$ is the electron temperature. Here, the electron number density $n_e$ and the electron temperature $T_e$ are calculated as the number of particles and thermal energy in a cell as

\begin{equation}
    T = \frac{m N}{3 k_B (N - 1)} \left(\frac{\sum_k v_k^2}{N} - u ^2\right),
\end{equation}

\noindent
where $N$ is the number of particles in a cell, $k$ is the index of the particles, $v_k$ is the velocity of the particle $k$, and $u$ is the flow velocity. They were calculated from the positions and velocities of the electrons. It should be noted that, although the isotropically averaged energy was used for the temperature calculation, the anisotropic effect would also have to be taken into account to investigate the plasma temperature in detail.

The net electron current density in the $z$-direction $j_{e,z}$ is calculated from the electron motions in the PIC-MCC simulation as

\begin{equation}
    \label{eq:current_density}
    j_{e,z} = - e n_e u_{e,z},
\end{equation}

\noindent
where $u_{e,z}$ is the electron flow velocity along the $z$-direction. It should be noted that the net electron current density in the $z$-direction $j_{e,z}$ includes both the $\vector{E} \times \vector{B}$ and diamagnetic effects, in addition to other drifts such as the grad-$B$ and curvature drift. In this simulation, the ion current density in the $z$-direction was neglected because it is sufficiently smaller than the electron current density.

We obtain the Lorentz force density in the $x$-direction $f_x$, which is exerted on the electrons and calculated as 

\begin{equation}
    \label{eq:lorentz_force_density}
    f_x = - j_{e,z} B_y.
\end{equation}

\noindent
Moreover, we obtain the net magnetic nozzle thrust $F_x$ by integrating the Lorentz force density in the $x$-direction $f_x$ in the downstream region as

\begin{equation}
    \label{eq:thrust}
    F_x = \int f_x dxdy.
\end{equation}

\section{RESULTS AND DISCUSSION}\label{results_and_discussion}

\begin{figure}
    \centering
    \includegraphics[width = \linewidth]{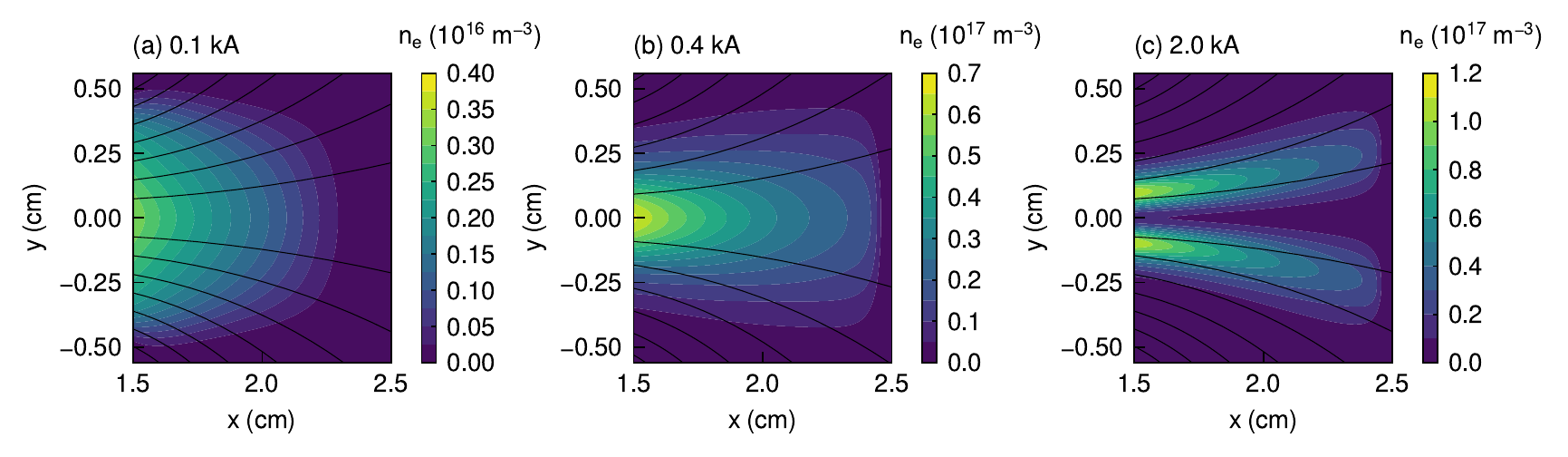}
    \caption{$x$-$y$ profiles of the electron number density $n_e$ for the three solenoid currents of (a) 0.1, (b) 0.4, and (c) 2.0 kA. The solid black curves represent the magnetic field lines produced by the solenoid.}
    \label{fig:electron_number_density}
\end{figure}

\begin{figure}
    \centering
    \includegraphics{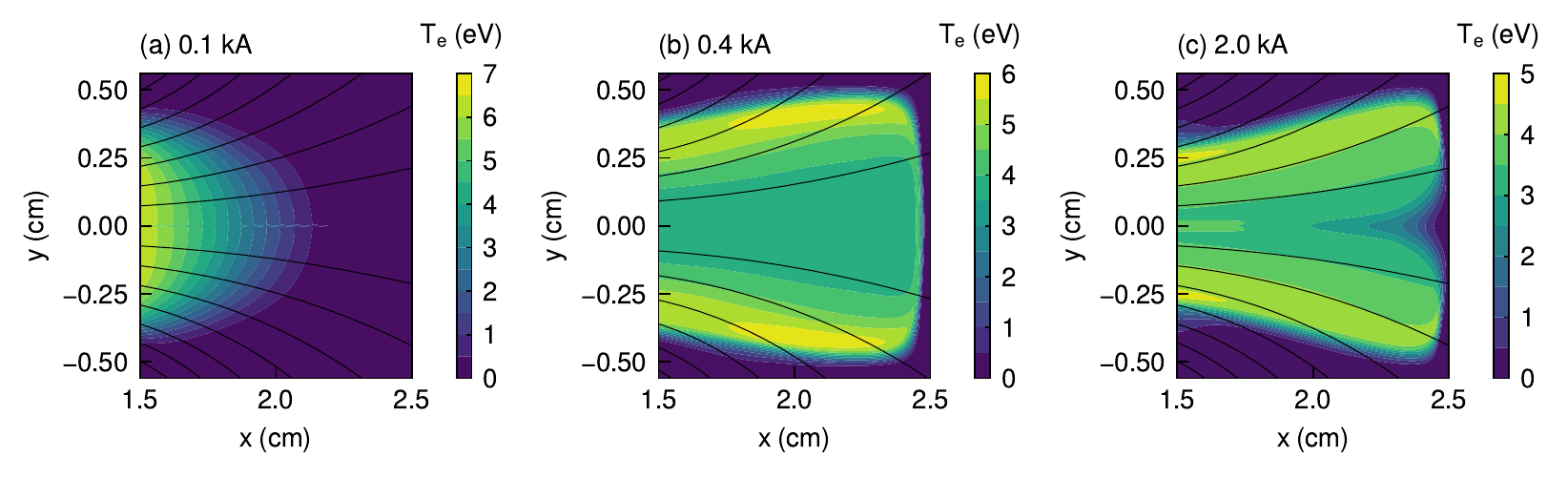}
    \caption{$x$-$y$ profiles of the electron temperature $T_e$ for the three solenoid currents of (a) 0.1, (b) 0.4, and (c) 2.0 kA. The solid black curves represent the magnetic field lines produced by the solenoid.}
    \label{fig:electron_temperature}
\end{figure}

\begin{figure}
    \centering
    \includegraphics[width = \linewidth]{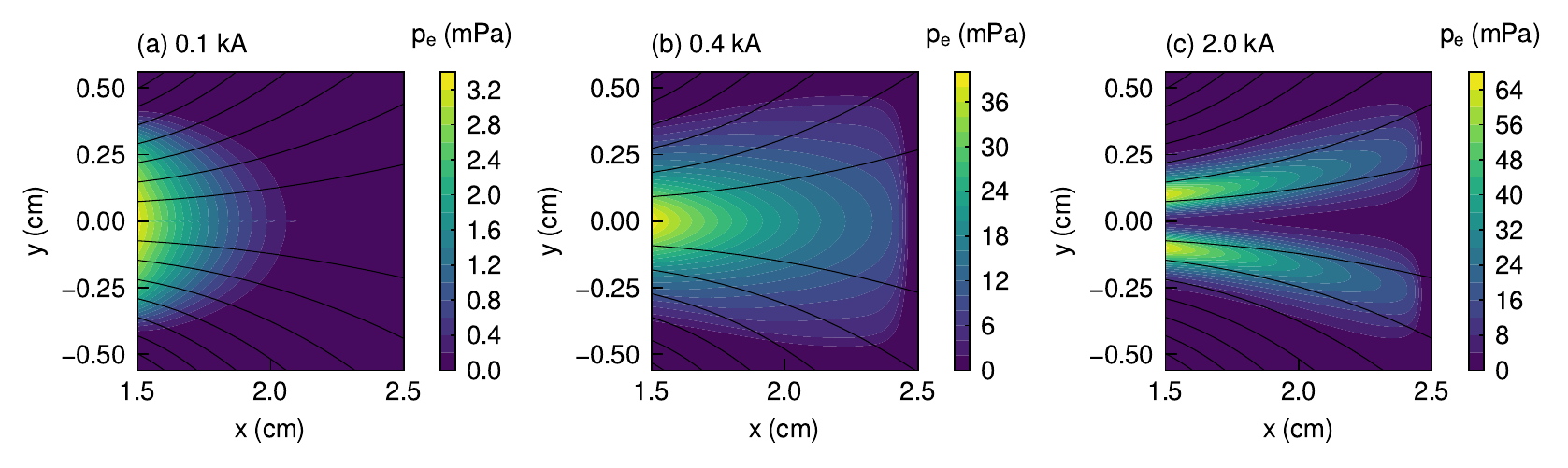}
    \caption{$x$-$y$ profiles of the electron pressure $p_e$ for the three solenoid currents of (a) 0.1, (b) 0.4, and (c) 2.0 kA. The solid black curves represent the magnetic field lines produced by the solenoid.}
    \label{fig:electron_pressure}
\end{figure}

Figures \ref{fig:electron_number_density} and \ref{fig:electron_temperature} show the $x$-$y$ profiles of the electron number density $n_e$ and the electron temperature $T_e$, respectively, for the three solenoid currents of 0.1, 0.4, and 2.0 kA. Then, the electron pressure $p_e$ is calculated from the electron number density $n_e$ in Fig. \ref{fig:electron_number_density} and the electron temperature $T_e$ in Fig. \ref{fig:electron_temperature}. Figure \ref{fig:electron_pressure} shows the $x$-$y$ profiles of the electron pressure $p_e$ for the three solenoid currents of 0.1, 0.4, and 2.0 kA. The $y$ profiles of the electron pressure $p_e$ for the solenoid currents of 0.1 and 0.4 kA indicate center-peaked plasmas as shown in Fig.~\ref{fig:electron_pressure}(a) and \ref{fig:electron_pressure}(b), but that for the solenoid current of 2.0 kA indicates a bimodal shape as shown in Fig.~\ref{fig:electron_pressure}(c). The reason for the bimodal shape is that energetic electrons are generated in a peripheral region near the rf antenna and the strong magnetic field prevents the plasma from being transported on the central $x$-axis, well reproducing the previous experimental results.\supercite{Takahashi2009_apl, Charles2010_apl, Ghosh2017_pop, Gulbrandsen2017_fr, Takahashi2017_pop} The maximum electron pressure increases as the solenoid current $I_B$ increases because the strong magnetic field confines the plasma more efficiently. A marked diamagnetic effect would be observed in regions where the electron pressure gradient $\nabla p_e$ is large. Here, because both the cylindrical thruster in experiments and the Cartesian one in simulations give rise to plasma with a bimodal shape, the cylindrical and Cartesian coordinates are considered not to have any essential differences.

\begin{figure}
    \centering
    \includegraphics{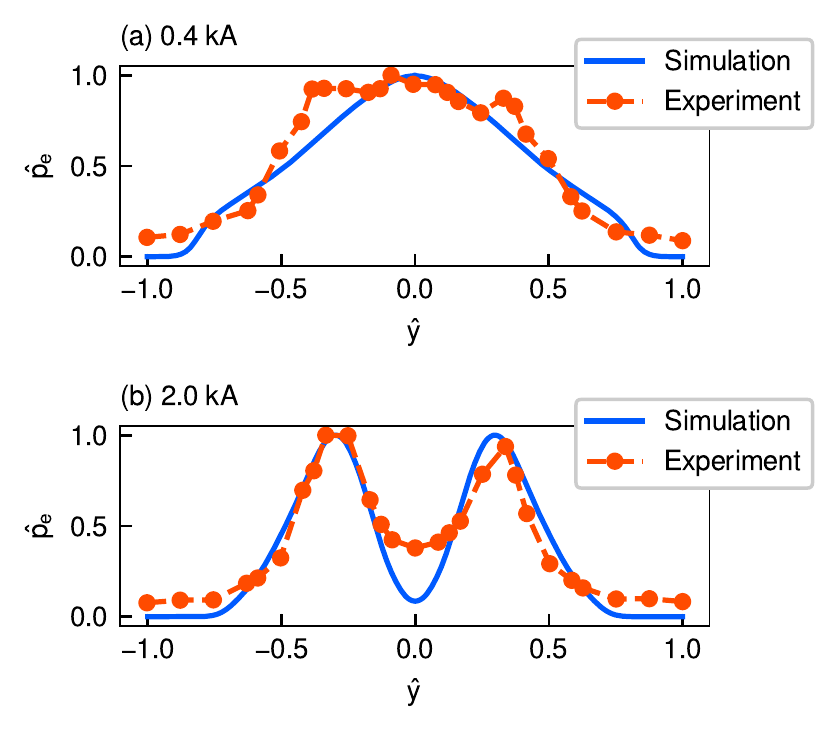}
    \caption{$\hat{y}$ profiles of the normalized electron pressure $\hat{p}_e$ for the two solenoid currents of (a) 0.4 and (b) 2.0 kA. The coordinate $y$ is normalized by the maximum position $y_{max}$ as $\hat{y} = y / y_{max}$, and the electron pressure $p_e$ is normalized by the maximum pressure $p_{e, max}$ as $\hat{p}_e = p_e / p_{e, max}$. The simulation result is a cross-section of $x$ = 2.0 cm at $I_B$ = 0.4 and 2.0 kA, and the experimental result is a cross-section of $z$ = 5 cm at $I_B$ = 2 and 12 A in Ref. \citenum{Takahashi2016_psst}.}
    \label{fig:section_electron_pressure}
\end{figure}

We compare the electron pressure $p_e$ computed with our PIC-MCC simulation with the experimental results from a previous study, which were measured using a Langmuir probe and is shown in Fig.~6(d) of Ref. \citenum{Takahashi2016_psst}. Figure \ref{fig:section_electron_pressure} shows the $\hat{y}$ profiles of the normalized electron pressure $\hat{p}_e$ for the two solenoid currents of 0.4 and 2.0 kA, where the coordinate $y$ is normalized by the maximum position $y_{max}$ as $\hat{y} = y / y_{max}$ and the electron pressure $p_e$ is normalized by the maximum pressure $p_{e, max}$ as $\hat{p}_e = p_e / p_{e, max}$. The electron pressure, $p_e$, in Fig.~\ref{fig:section_electron_pressure} was calculated using Eq. \eqref{eq:electron_pressure}, where $n_e$ and $T_e$ in the experiments are obtained from the current-voltage characteristics assuming a Maxwellian electron energy distribution function. Here, the simulation result is the cross-section of $x$ = 2.0 cm at $I_B$ = 0.4 and 2.0 kA, and the experimental result is that of $z$ = 5 cm at $I_B$ = 2 and 12 A, respectively. The position of $x$ = 2.0 cm in the simulation roughly corresponds to that of $z$ = 5 cm in the experiment because the size of the calculation model is approximately one-sixth as described in Sec.~\ref{model}. The normalized electron pressure $\hat{p}_e$ obtained from our simulation is in good agreement with the experimental results. The simulation results shown in Fig.~\ref{fig:section_electron_pressure}(b) indicates the low pressure at the center of $\hat{y}$ = 0 compared with the experimental result because the calculation model is two-dimensional ($x$-$y$), whereas actual thrusters have a cylindrical shape. The cylindrical discharge chamber confines the plasma to the radial direction. However, the two-dimensional model with its rectangular coordinates is less capable of structurally confining the plasma to the center because free vertical motions exist in the $z$-direction. The electron pressure gradient $\nabla p_e$ becomes large compared with the experimental result; thus, our two-dimensional model slightly overestimates the diamagnetic effects in the central region. However, the normalized electron pressure $\hat{p}_e$ obtained from our simulation is qualitatively consistent with the experimental result, indicating a confirmation of the validity of the simulation results, although it may not quantitatively capture the physics of cylindrical thrusters. 

In these simulations, the ratio $r_L/L$ is roughly adjusted between the simulations and the experiments, where $r_L$ is the Larmor radius of ions, $L$ is the characteristic length, and $L$ is set to the thruster height (diameter in cylindrical coordinates). However, the parameters including the ratio $r_L/L$ are not scaled up or down accurately in the simulations. Instead of scaling, we consider the distribution of the electron pressure to be a key parameter for validating the experimental results. If the plasma pressure distributions are the same, the simulations are expected to reproduce the internal plasma current and force quantitatively. In addition to the plasma profiles, the Larmor radii were roughly the same between the simulations and the experiments, and the Hall parameters were in excess of a few hundred in both cases, indicating that there were no essential differences between the simulations and the experiments.

\begin{figure}
    \centering
    \includegraphics{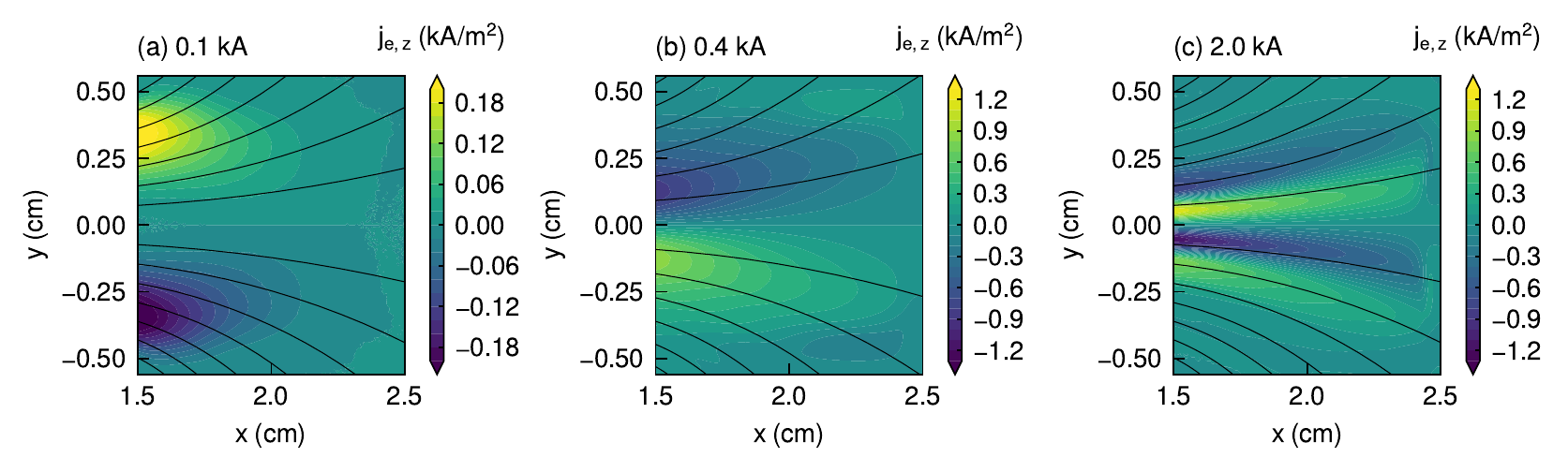}
    \caption{$x$-$y$ profiles of the net electron current density $j_{e,z}$ for the three solenoid currents of (a) 0.1, (b) 0.4, and (c) 2.0 kA.}
    \label{fig:electron_current_density_3}
\end{figure}

Figure \ref{fig:electron_current_density_3} shows the $x$-$y$ profiles of the net electron current density $j_{e,z}$ for the three solenoid currents of 0.1, 0.4, and 2.0 kA, where the net electron current density $j_{e,z}$ is calculated using Eq.~\eqref{eq:current_density}. The net electron current densities in the upper and lower half regions are opposite to each other in Figs.~\ref{fig:electron_current_density_3}(a) and \ref{fig:electron_current_density_3}(b). However, in Fig.~\ref{fig:electron_current_density_3}(c), the opposite net electron current density $j_{e,z}$ only exists in the upper half region. The reason for this unusual phenomenon is mentioned later in this paper. Although the net electron current density does not follow the magnetic field lines in the case of the weak magnetic field strength as shown in Fig.~\ref{fig:electron_current_density_3}(a), they are fairly aligned with the magnetic field lines for $I_B$ = 0.4 kA as shown in Fig.~\ref{fig:electron_current_density_3}(b). Figure~\ref{fig:electron_current_density_3}(c) shows that the net electron current density is clearly distributed along the magnetic field lines because of the strong magnetic field. 

Here, we define the characteristic magnetic field strength as the magnetic field strength under the solenoid at $x$ = 1.1 cm and on the $x$-axis. For solenoid currents of 0.1, 0.4, and 2.0 kA, the characteristic magnetic field strengths are approximately 5, 20, and 100 mT, respectively. Then, the respective Larmor radii become 1.70, 0.43, and 0.08 mm at the electron temperature of 5 eV. In addition, the Hall parameters were estimated to be more than one hundred for any solenoid current. Therefore, electrons in the magnetic nozzle are not likely to be transported across the magnetic field lines by elastic collisions. Moreover, because the Larmor radius for the solenoid current of 2.0 kA is sufficiently small compared with the thruster size, the plasma profile is split into the two regions.

\begin{figure}
    \centering
    \includegraphics{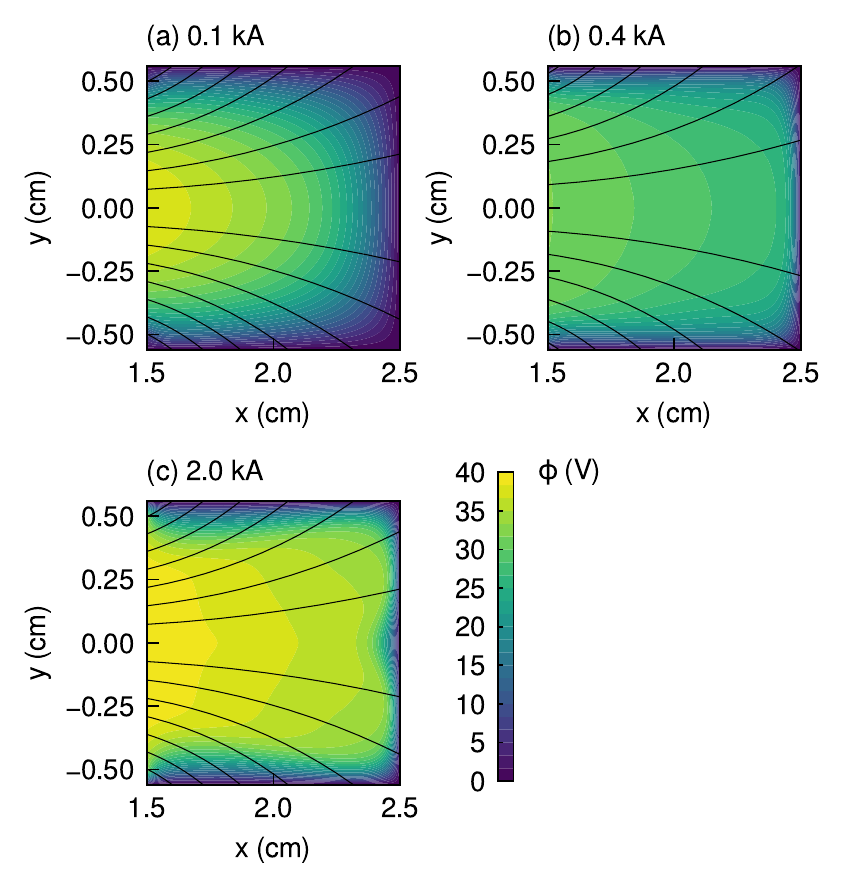}
    \caption{$x$-$y$ profiles of the potential $\phi$ for the three solenoid currents of (a) 0.1, (b) 0.4, and (c) 2.0 kA. The solid black curves represent the magnetic field lines produced by the solenoid.}
    \label{fig:potential}
\end{figure}

We derived the potential distributions to investigate the $\vector{E} \times \vector{B}$ effect. Figure \ref{fig:potential} shows the $x$-$y$ profiles of the potential $\phi$ for the three solenoid currents of 0.1, 0.4, and 2.0 kA. The potential decreases from the plasma source on the left side to the downstream on the right side at all solenoid currents. The potential gradient $\nabla \phi$ and the solenoid magnetic field $\vector{B}$ would be expected to induce an $\vector{E} \times \vector{B}$ drift current. It should be noted that there are sheaths near the right boundary at $x$ = 2.3--2.5 cm and the top and bottom boundaries at $y$ = $\pm$(0.3--0.56) cm because the potential $\phi$ is solved with the Dirichlet boundary condition as $\phi$ = 0.

\begin{figure}
    \centering
    \includegraphics[width = \linewidth]{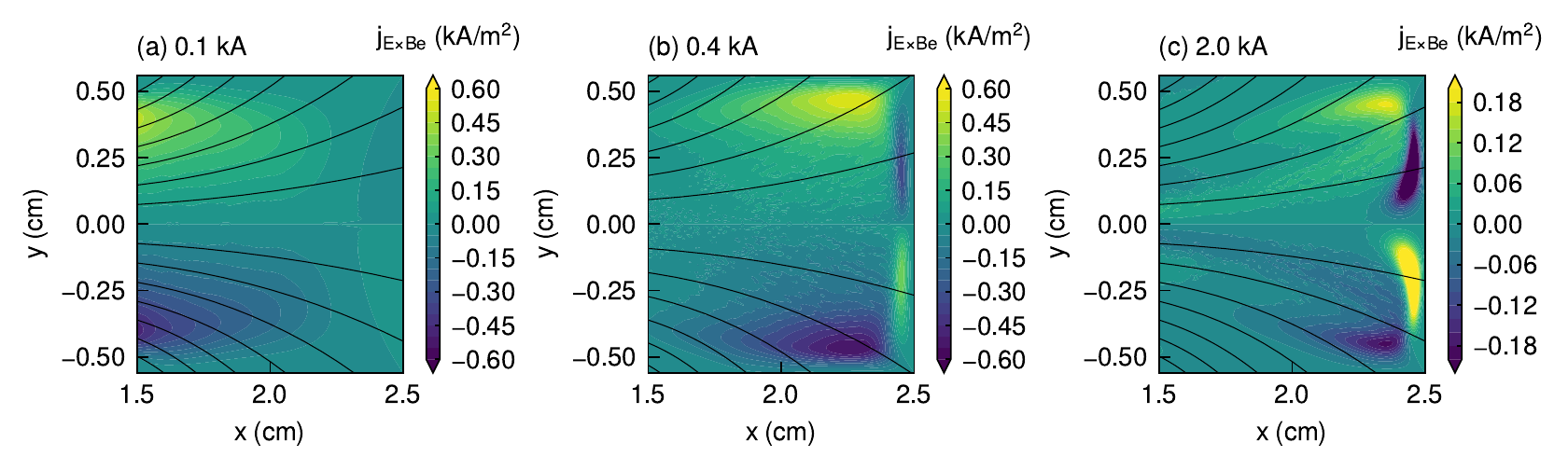}
    \caption{$x$-$y$ profiles of the $\vector{E} \times \vector{B}$ drift current density $j_{E \times Be}$ for the three solenoid currents of (a) 0.1, (b) 0.4, and (c) 2.0 kA.}
    \label{fig:exb_current_density}
\end{figure}

\begin{figure}
    \centering
    \includegraphics[width = \linewidth]{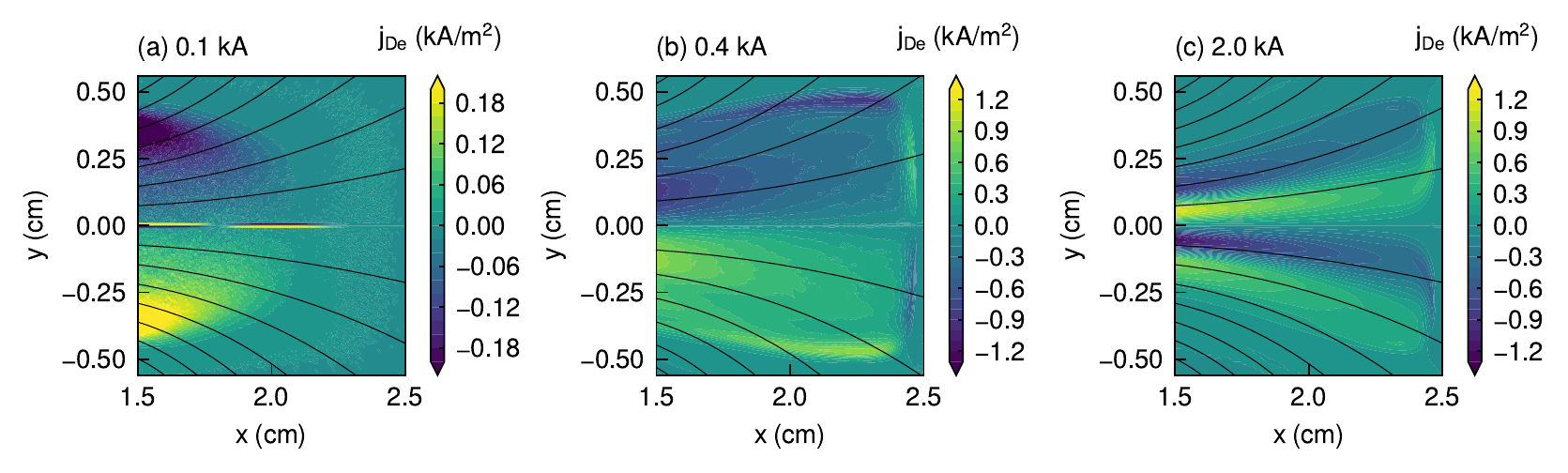}
    \caption{$x$-$y$ profiles of the diamagnetic drift current density $j_{De}$ for the three solenoid currents of (a) 0.1, (b) 0.4, and (c) 2.0 kA.}
    \label{fig:electron_diamagnetic_current_density}
\end{figure}

We calculate the $\vector{E} \times \vector{B}$ drift current density $j_{E \times Be}$ and the diamagnetic drift current density $j_{De}$ for comparison with the net electron current density $j_{e,z}$, which are expressed in terms of the two-dimensional fluid model as

\begin{equation}
    \label{eq:exb}
    j_{E \times Be} = - e n_e \frac{E_{es,x} B_y - E_{es,y} B_x}{B^2},
\end{equation}

\begin{equation}
    \label{eq:diamagnetic_current}
    j_{De} = - \frac{\partial p_e}{\partial x} \frac{B_y}{B^2} + \frac{\partial p_e}{\partial y} \frac{B_x}{B^2},
\end{equation}

\noindent
respectively. Figures \ref{fig:exb_current_density} and \ref{fig:electron_diamagnetic_current_density} show the $x$-$y$ profiles of the $\vector{E} \times \vector{B}$ drift current density $j_{E \times Be}$ and the diamagnetic drift current density $j_{De}$ for the three solenoid currents of 0.1, 0.4, and 2.0 kA, respectively. Here, we focus on the region of 1.5 cm < $x$ < 2.3 cm and $|y|$ < 0.3 cm, where the sheath does not affect the plasma. The $\vector{E} \times \vector{B}$ and diamagnetic current densities in the upper and lower half regions are opposite to each other, similar to the net electron current density in Fig.~\ref{fig:electron_current_density_3}). However, Fig.~\ref{fig:electron_diamagnetic_current_density}(c) shows that the diamagnetic current density $j_{De}$ exists in the opposite direction to itself only in the upper half region because of the bimodal shape of the electron pressure $p_e$ in Fig.~\ref{fig:electron_pressure}(c).

\begin{figure}
    \centering
    \includegraphics{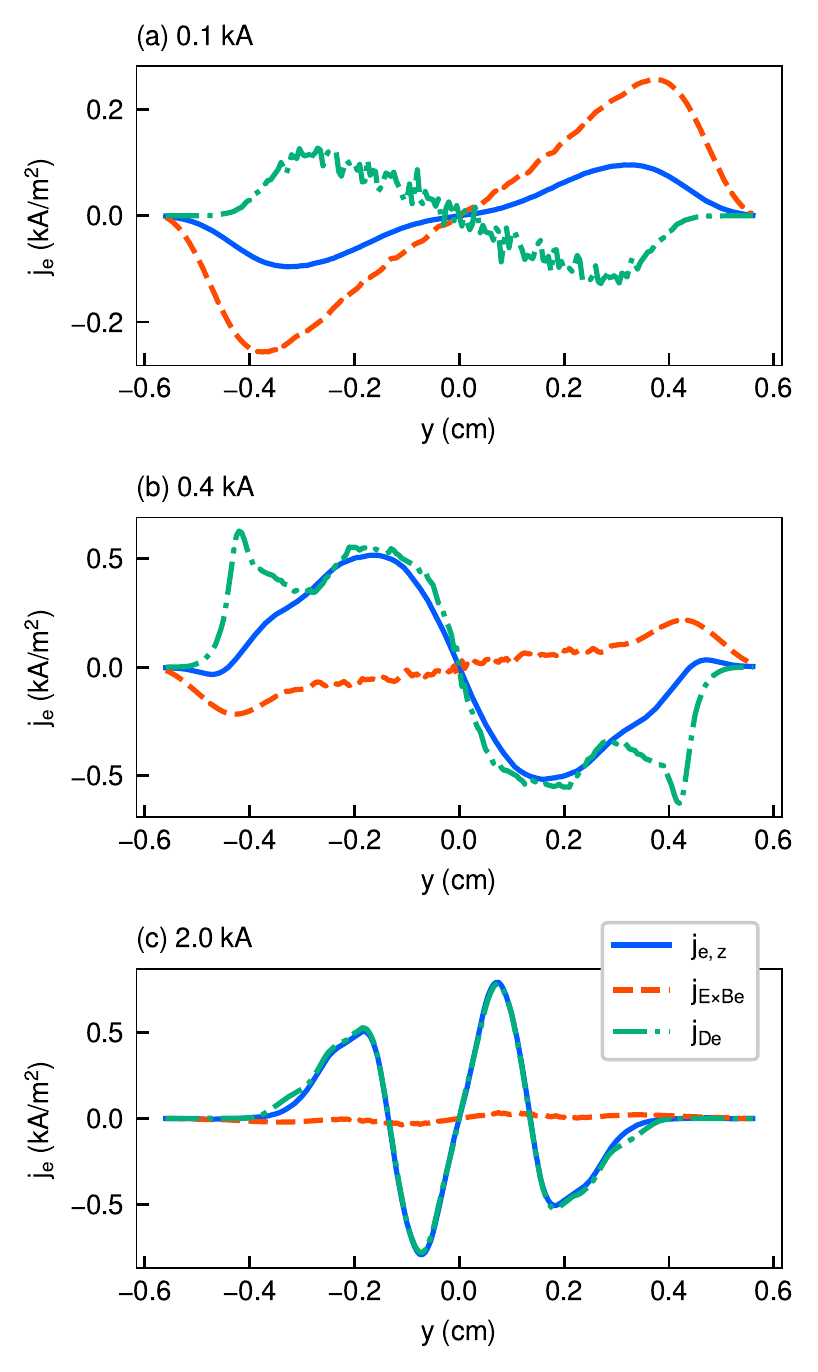}
    \caption{$y$-profiles of the net electron current density $j_{e,z}$ (solid blue curve), the $\vector{E} \times \vector{B}$ drift current density $j_{E \times Be}$ (dashed orange curve), and the diamagnetic drift current density $j_{De}$ (dotted-dashed green curve) at $x$ = 1.8 cm for the three solenoid currents of (a) 0.1, (b) 0.4, and (c) 2.0 kA.}
    \label{fig:section_electron_current_density_3}
\end{figure}

Figure \ref{fig:section_electron_current_density_3} shows the $y$ profiles of the net electron current density $j_{e,z}$, the $\vector{E} \times \vector{B}$ drift current density $j_{E \times Be}$, and the diamagnetic drift current density $j_{De}$ at $x$ = 1.8 cm for the three solenoid currents of 0.1, 0.4, and 2.0 kA. Although both the $\vector{E} \times \vector{B}$ and the diamagnetic effects exist in all the cases, they are canceled out because they exist in opposite directions to each other, as is obvious from the net electron current density $j_{e,z}$. In the case of the solenoid current of 0.1 kA, the $\vector{E} \times \vector{B}$ and diamagnetic current densities are between $\pm$0.25 kA/m$^2$, which is small compared with those of 0.4 and 2.0 kA. Although the $\vector{E} \times \vector{B}$ drift current density $j_{E \times Be}$ is larger than the diamagnetic one $j_{De}$ at $I_B = 0.1$ kA, the diamagnetic drift current density $j_{De}$ is larger than the $\vector{E} \times \vector{B}$ one $j_{E \times Be}$ for both the solenoid currents of 0.4 and 2.0 kA. The effect of the $\vector{E} \times \vector{B}$ drift current density $j_{E \times Be}$ decreases as the solenoid current increases, and the diamagnetic drift current density $j_{De}$ becomes dominant. In addition, other drifts such as grad-$B$ and curvature drifts are negligibly small in the presence of the strong magnetic field because the net electron current density $j_{e,z}$ is almost equal to the diamagnetic current density $j_{De}$ as shown in Fig.~\ref{fig:section_electron_current_density_3}(c). 

The Dirichlet boundary condition at $x$ = 2.5 cm and $y$ = $\pm$0.56 cm, which imposes a zero potential, affects the simulation results. Here, a marked sheath effect is expected to be seen in the $\vector{E} \times \vector{B}$ current in Fig.~\ref{fig:exb_current_density}, because it is directly determined by the electrostatic field $\vector{E}_{es}$. In Fig.~\ref{fig:exb_current_density}, anomalous current densities exist in the regions $x$ = 2.3--2.5 cm and $|y|$ > 0.3 cm, and these are not suitable for validating the experimental results. However, anomalous current densities do not exist in $x$ = 1.5--2.3 cm and $|y|$ < 0.3 cm, indicating that the sheath effect is sufficiently small in these regions. Because the $\vector{E} \times \vector{B}$ current density is larger when the solenoid current is 0.1 kA as shown in Fig.~\ref{fig:section_electron_current_density_3}(a), the sheath effect may be relatively large. However, when the solenoid current is 2.0 kA, the $\vector{E} \times \vector{B}$ current density is negligibly small, as shown in Fig.~\ref{fig:section_electron_current_density_3}(c), and the sheath does not affect the result, i.e., the diamagnetic effect dominates the net current density.

The $\vector{E} \times \vector{B}$ drift current density $j_{E \times Be}$ is proportional to the electrostatic field $E_{es}$ and inversely proportional to the magnetic field strength $B$ as indicated in Eq.~\eqref{eq:exb}: The electrostatic field $\vector{E}_{es}$ tends to remain unchanged with increasing solenoid current $I_B$, as shown in Fig.~\ref{fig:potential}, and thus the $\vector{E} \times \vector{B}$ drift current density $j_{E \times Be}$ decreases as the solenoid current $I_B$ increases. 

The diamagnetic drift current density $j_{De}$ is proportional to the electron pressure gradient $\nabla p_e$ and inversely proportional to the solenoid magnetic field $B$, as indicated in Eq.~\eqref{eq:diamagnetic_current}. The electron pressure gradient $\nabla p_e$ increases significantly as the solenoid current increases from 0.1 to 0.4 kA because the maximum electron pressure increases tenfold as shown in Figs.~\ref{fig:electron_pressure}(a) and \ref{fig:electron_pressure}(b). This higher rate of increase in $\nabla p_e$ than $B$ results in a more than two-fold increase in $j_{De}$. However, the increase in the maximum electron pressure from 0.4 to 2.0 kA is only twofold, and the plasma width becomes roughly half in the $y$-direction because of the bimodal shape, approximately quadrupling the electron pressure gradient. Thus, the increasing rate of $\nabla p_e$ is approximately equal to that of $B$, leading to a slight increase in $j_{De}$. 

The net electron current density $j_{e,z}$ is dominated by the diamagnetic drift current density $j_{De}$ at $I_B$ = 2.0 kA as shown in Fig.~\ref{fig:section_electron_current_density_3}(c). A similar trend was observed in a previous experiment (see Figs.~6(e) and 6(f) in Ref.~\citenum{Takahashi2016_psst}). Our simulation results were in good agreement with the experimental results.

\begin{figure}
    \centering
    \includegraphics{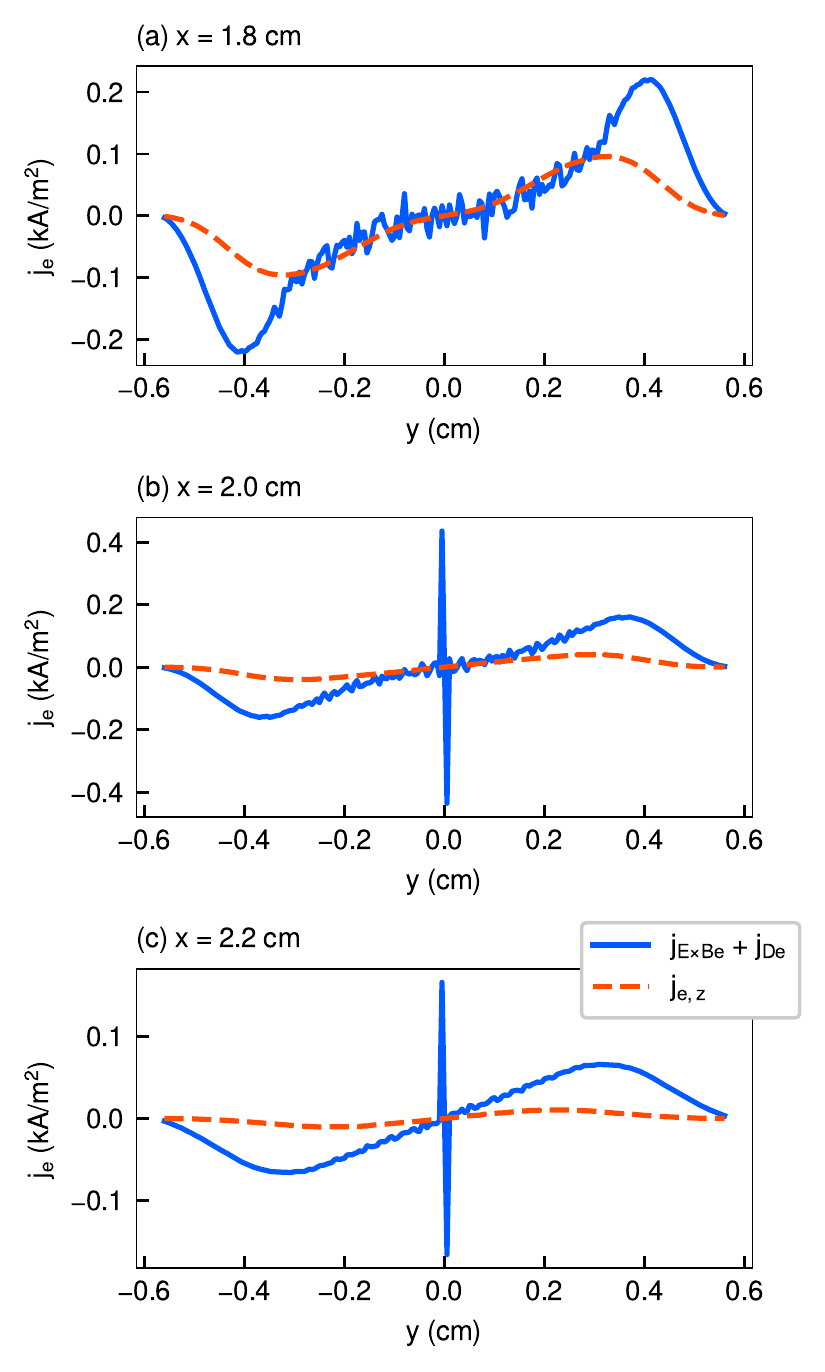}
    \caption{$y$ profiles of the sum of the $\vector{E} \times \vector{B}$ drift current density and the diamagnetic drift current density $j_{E \times Be} + j_{De}$ (solid blue curve) and the net electron current density $j_{e,z}$ (dashed orange curve) at $x$ = (a) 1.8, (b) 2.0, and (c) 2.2 cm for the solenoid current of 0.1 kA.}
    \label{fig:section_electron_current_density_3_01}
\end{figure}

\begin{figure}
    \centering
    \includegraphics{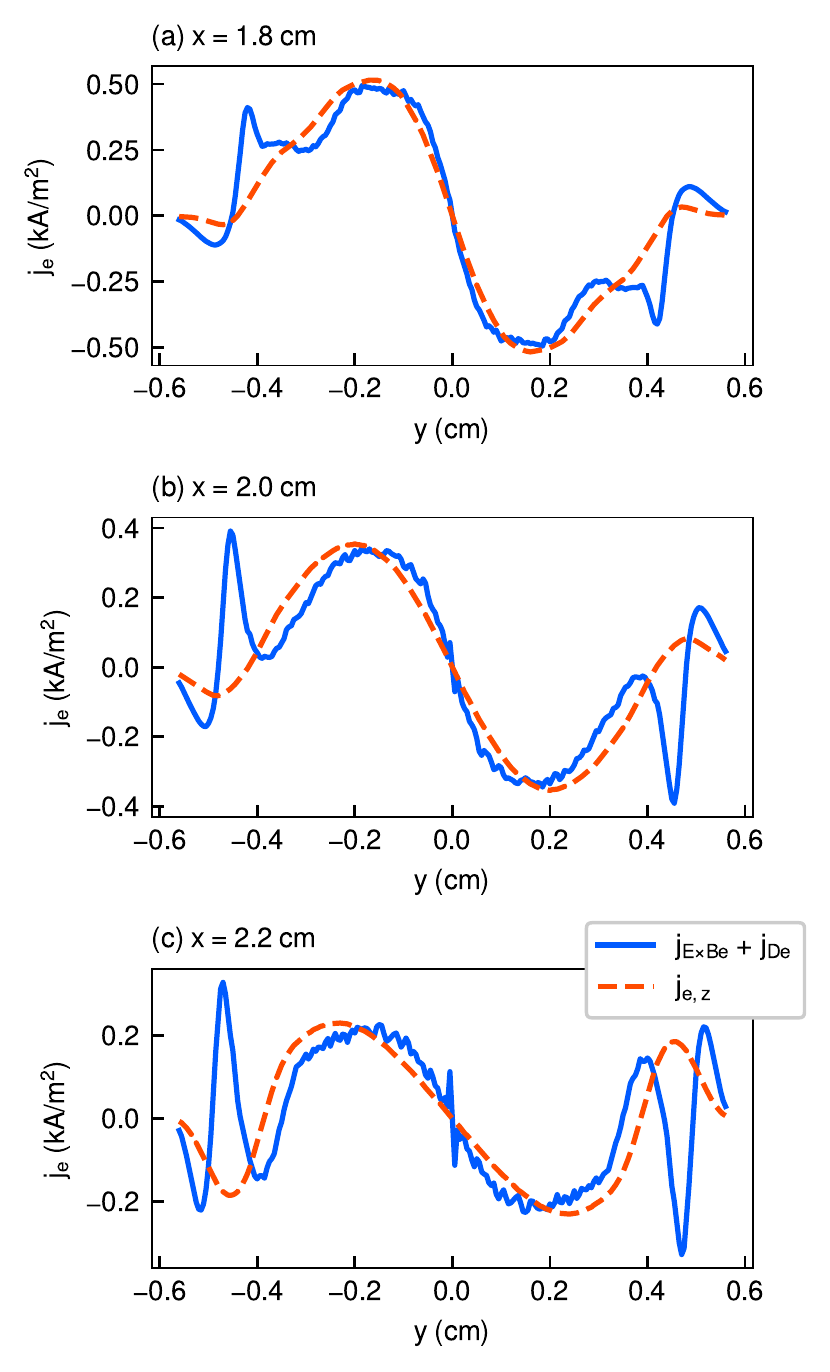}
    \caption{$y$ profiles of the sum of the $\vector{E} \times \vector{B}$ drift current density and the diamagnetic drift current density $j_{E \times Be} + j_{De}$ (solid blue curve) and the net electron current density $j_{e,z}$ (dashed orange curve) at $x$ = (a) 1.8, (b) 2.0, and (c) 2.2 cm for the solenoid current of 0.4 kA.}
    \label{fig:section_electron_current_density_3_04}
\end{figure}

\begin{figure}
    \centering
    \includegraphics{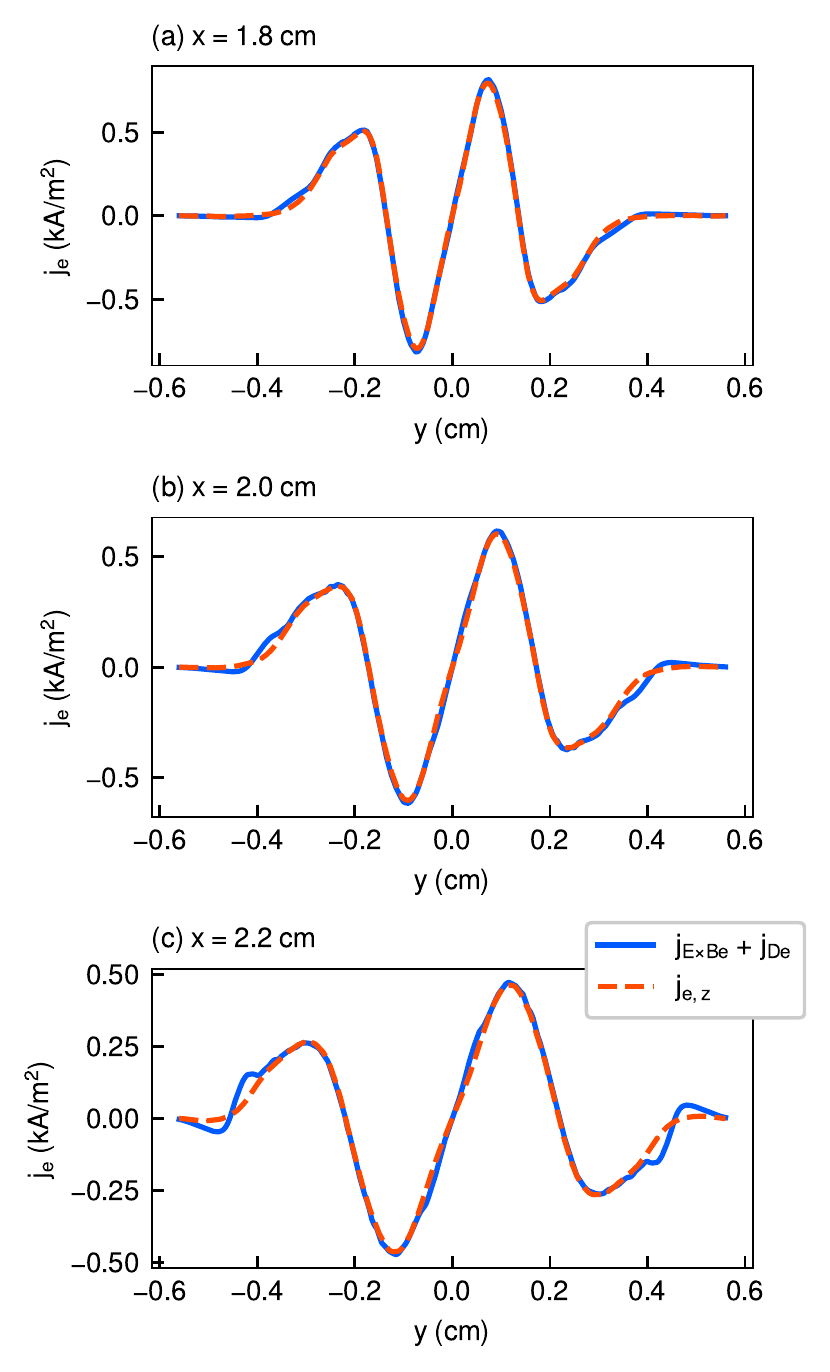}
    \caption{$y$ profiles of the sum of the $\vector{E} \times \vector{B}$ drift current density and the diamagnetic drift current density $j_{E \times Be} + j_{De}$ (solid blue curve) and the net electron current density $j_{e,z}$ (dashed orange curve) at $x$ = (a) 1.8, (b) 2.0, and (c) 2.2 cm for the solenoid current of 2.0 kA.}
    \label{fig:section_electron_current_density_3_20}
\end{figure}

Figures \ref{fig:section_electron_current_density_3_01}--\ref{fig:section_electron_current_density_3_20} show the $y$-profiles of the sum of the $\vector{E} \times \vector{B}$ and diamagnetic drift current density $j_{E \times Be} + j_{De}$ and the net electron current density $j_{e,z}$ at $x$ = 1.8, 2.0, and 2.2 cm for the three solenoid currents of 0.1, 0.4, and 2.0 kA. Note that $j_{E \times Be}$ + $j_{De}$ for the solenoid current of 0.1 kA at $x$ = 2.0 and 2.2 cm include anomalous results near $y$ = 0 cm. For the solenoid current of 0.1 kA, $j_{E \times Be} + j_{De}$ is equal to $j_{e,z}$ at $x$ = 1.8 cm for $|y|$ < 0.3 cm, but they do not correspond at $x$ = 2.0 and 2.2 cm. Therefore, $j_{e,z}$ is composed of not only $j_{E \times Be}$ and $j_{De}$ but also other drift currents, e.g., grad-$B$ and curvature drifts. For the solenoid current of 0.4 kA, $j_{E \times Be} + j_{De}$ is roughly equal to $j_{e,z}$ at all locations. In addition, $j_{E \times Be} + j_{De}$ is completely equal to $j_{e,z}$ for the solenoid current of 2.0 kA except in the peripheral region within |$y$| > 0.3 cm, where the sheath is expected to affect the plasma. Because the $\vector{E} \times \vector{B}$ effect decreases as the solenoid current increases as shown in Fig.~\ref{fig:section_electron_current_density_3}, increasing the solenoid current suppresses the $\vector{E} \times \vector{B}$ effect and other effects such as grad-$B$ and curvature drifts, and the diamagnetic effect becomes dominant instead.

\begin{figure}
    \centering
    \includegraphics{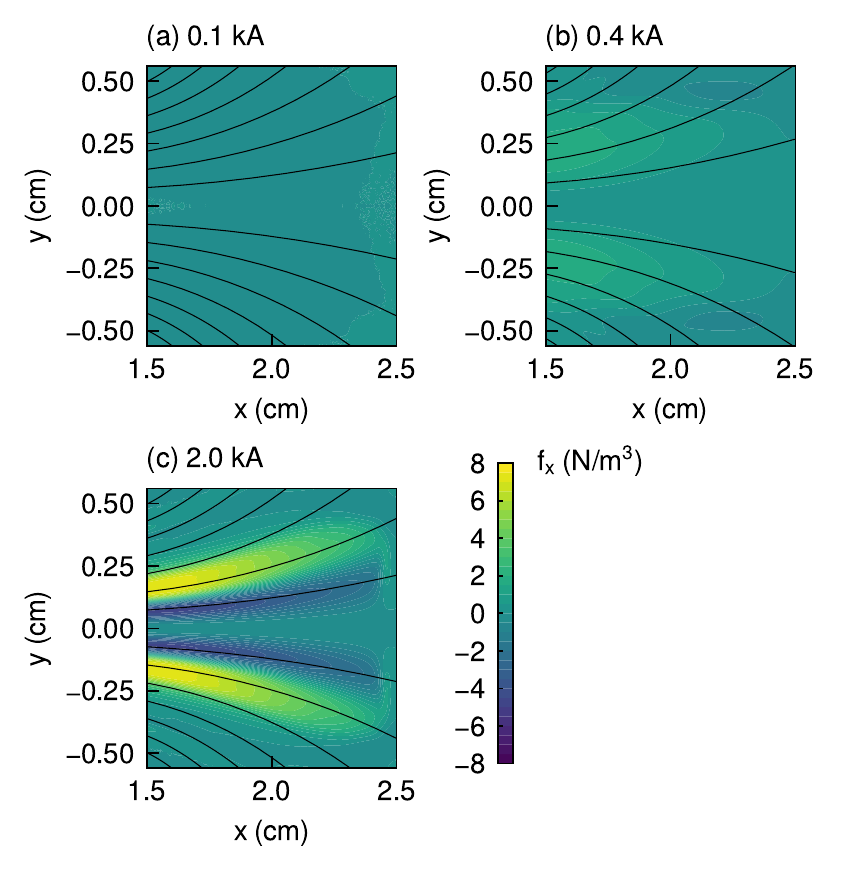}
    \caption{$x$-$y$ profiles of the Lorentz force density in the $x$-direction $f_x$ exerted on electrons for the three solenoid currents of (a) 0.1, (b) 0.4, and (c) 2.0 kA. The Lorentz force density in the $x$-direction $f_x$ is calculated from the electron current density $j_{e,z}$ and the solenoid magnetic field in the $y$-direction $B_y$. The solid black curves represent the magnetic field lines produced by the solenoid.}
    \label{fig:lorentz_force_density}
\end{figure}

We obtain the Lorentz force density exerted on the electrons in the $x$-direction ($f_x$), which is calculated from the net electron current density $j_{e,z}$ and the solenoid magnetic field in the $y$-direction $B_y$, as shown in Eq.~\eqref{eq:lorentz_force_density}. Figure \ref{fig:lorentz_force_density} shows the $x$-$y$ profiles of the Lorentz force density in the $x$-direction ($f_x$) exerted on the electrons for the three solenoid currents of 0.1, 0.4, and 2.0 kA. The Lorentz force density in the $x$-direction ($f_x$) increases dramatically with an increase in the solenoid current $I_B$. However, the magnitude of the net electron current density $j_{e,z}$ does not change significantly when the solenoid current $I_B$ increases, especially in the case of the solenoid currents of 0.4 and 2.0 kA as shown in Figs.~\ref{fig:section_electron_current_density_3}(b) and \ref{fig:section_electron_current_density_3}(c). Here, the Lorentz force density in the $x$-direction ($f_x$) is the product of the net electron current density $j_{e,z}$ and the solenoid magnetic field in the $y$-direction $B_y$, as shown in Eq.~\eqref{eq:lorentz_force_density}; additionally, the magnetic field strength $B$ of the solenoid increases with an increase in the solenoid current $I_B$. Therefore, the strong magnetic field pushes electrons downstream electromagnetically, although the electron current density does not change significantly with the solenoid current $I_B$.

\begin{figure}
    \centering
    \includegraphics{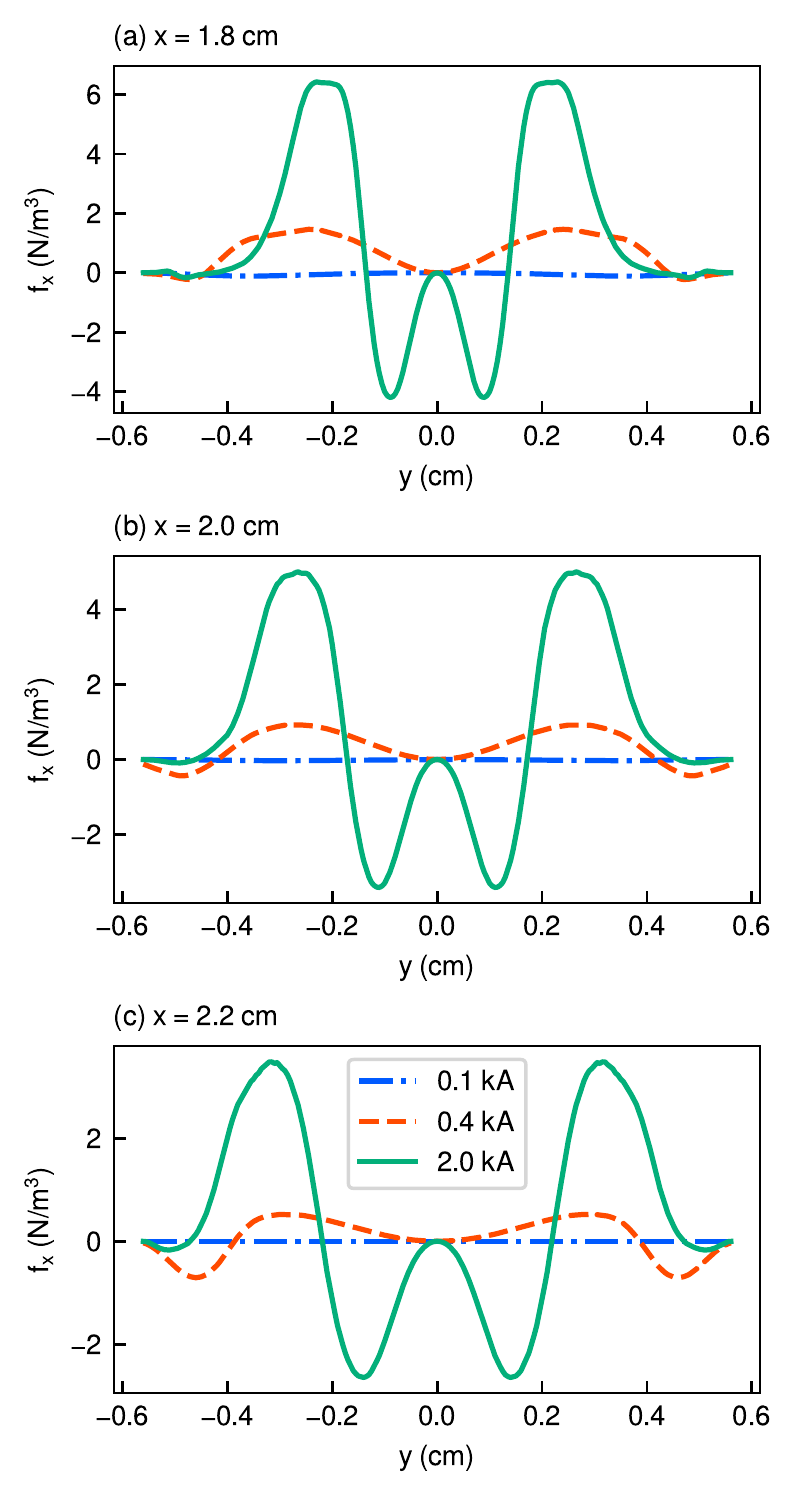}
    \caption{$y$ profiles of the Lorentz force density in the $x$-direction $f_x$ at $x$ = (a) 1.8, (b) 2.0, and (c) 2.2 cm for the three solenoid currents of 0.1 (dotted-dashed blue curve), 0.4 (dashed orange curve), and 2.0 kA (solid green curve). The Lorentz force density in the $x$-direction $f_x$ is calculated from the electron current density $j_{e,z}$ and the magnetic field of the solenoid in the $y$-direction $B_y$.}
    \label{fig:section_lorentz_force_density_1}
\end{figure}

Figure \ref{fig:section_lorentz_force_density_1} shows the $y$ profiles of the Lorentz force density in the $x$-direction ($f_x$) exerted on electrons at $x$ = 1.8 cm, 2.0 cm, and 2.2 cm for the three solenoid currents of 0.1, 0.4, and 2.0 kA. The Lorentz force density in the $x$-direction ($f_x$) is almost zero for the solenoid current $I_B$ = 0.1 kA at all locations, whereas a positive Lorentz force density is obtained for the solenoid current of $I_B$ = 0.4 kA. In the case of $x$ = 1.8 cm and the solenoid current of 2.0 kA, the positive Lorentz force density in the $x$-direction is 6.43 N/m$^3$ at maximum, whereas the negative Lorentz force density in the $x$-direction is exerted on the electrons between $y$ = $\pm$ 0.1 cm. Therefore, the positive Lorentz force density in the $x$-direction increases as the solenoid current $I_B$ increases, whereas the negative density emerges in the central region.

Table \ref{tab:magnetic_nozzle_thrust} shows the magnetic nozzle thrust $F_x$ calculated by Eq.~\eqref{eq:thrust}, where the Lorentz force density in the $x$-direction $f_x$ is integrated in 1.5 cm < $x$ < 2.3 cm and $|y|$ < 0.3 cm to reduce the sheath effects. The magnetic nozzle thrust $F_x$ increases as the solenoid current $I_B$ increases. At the solenoid current $I_B$ = 0.1 kA, the magnetic nozzle thrust $F_x$ is $-$1.22 $\mu$N/m, indicating that the magnetic nozzle does not accelerate the plasma in the downstream direction. The negative magnetic nozzle thrust is due to the $\vector{E} \times \vector{B}$ drift current because it occurs in the direction opposite to the diamagnetic current and $- j_{E \times Be} B_y$ is in the negative $x$-direction. Note that the total thrust is not negative because the ions accelerated by the electrostatic force impart a positive thrust and the magnetic nozzle thrust is composed of only electron momentum. At the solenoid currents of 0.4 and 2.0 kA, positive magnetic nozzle thrusts are obtained. In addition, the magnetic nozzle thrust at $I_B$ = 2.0 kA is about 1.41 times larger than that at $I_B$ = 0.4 kA. Therefore, the strong magnetic field generates larger thrust, even though the thrust is also produced in the negative thrust region. The net electron current density $j_{e,z}$ remains almost unchanged as the solenoid current increases, suggesting that the strong magnetic field would increase the magnetic nozzle thrust $F_x$. 

The magnetic nozzle thrust $F_x$ does not depend directly on the mass ejection, as indicated in the Tsiolkovsky equation. In the magnetic nozzle, the electron pressure is converted to ion momentum by electrostatic fields such as a current-free double layer or an ambipolar electric field, and the thrust is finally obtained by the ion ejection \supercite{Fruchtman2006_prl}. However, the conversion of momentum between ions and electrons has not been clarified in detail and would have to be investigated in the future.

\begin{table}
    \centering
    \caption{Magnetic nozzle thrust $F_x$.}
    \label{tab:magnetic_nozzle_thrust}
    \begin{tabular}{ll}
        \hline
        $I_B$ (kA) & $F_x$ ($\mu$N/m) \\
        \hline
        0.1 & $-$1.22 \\
        0.4 & 34.2 \\
        2.0 & 48.3 \\
        \hline
    \end{tabular}
\end{table}

\section{CONCLUSION}

We conducted two-dimensional PIC-MCC simulations of an electrodeless plasma thruster using a magnetic nozzle for three solenoid currents and investigated the dependence of the internal plasma currents on the magnetic field strength. The diamagnetic drift current density becomes dominant with increasing solenoid current (i.e., magnetic field strength), whereas the $\vector{E} \times \vector{B}$ drift current density decreases, indicating that the net electron current is caused due to the diamagnetic effect. Other drifts such as grad-$B$ and curvature drifts are observed for a solenoid current of 0.1 kA in the simulation, but their effects diminish as the solenoid current increases. These results are consistent with the previous experimental results, which implied that the dominant drift changed from the $\vector{E} \times \vector{B}$ effect to the diamagnetic one as the magnetic field became stronger. The Lorentz force density has positive and negative regions at the highest solenoid current that was examined ($I_B$ = 2.0 kA), but the net magnetic nozzle thrust is positive and increases with increasing solenoid current. Therefore, our simulations demonstrated that the diamagnetic effect significantly contributes to thrust generation in the magnetic nozzle. Because the simulations with the sheath under the Dirichlet boundary condition reproduced the internal plasma currents measured in previous experiments, the sheath effect is not expected to be essential for the diamagnetic effect and thrust generation in the magnetic nozzle. However, the Dirichlet boundary condition is not suitable for validating the plasma in space operations. In the future, it would also be necessary to validate the internal plasma currents for operations in space with open boundary conditions, for example, the Neumann boundary condition.

\section*{ACKNOWLEDGEMENT}

This work was partly supported by JSPS KAKENHI Grant Numbers JP21J15345 and JP19H00663. The computer simulation was performed on the A-KDK computer system at the Research Institute for Sustainable Humanosphere, Kyoto University. One of the authors (K.E.) received a scholarship from the Futaba Foundation.

\section*{DATA AVAILABILITY}

The data that support the findings of this study are available from the corresponding author upon reasonable request.

\printbibliography

\end{document}